\documentclass[pdflatex,sn-nature,iicol]{sn-jnl}% Style for submissions to Nature Portfolio journals
\usepackage{graphicx}%
\usepackage{multirow}%
\usepackage{amsmath,amssymb,amsfonts}%
\usepackage{amsthm}%
\usepackage{mathrsfs}%
\usepackage{mathtools}
\usepackage[title]{appendix}%
\usepackage{xcolor}%
\usepackage{textcomp}%
\usepackage{manyfoot}%
\usepackage{booktabs}%
\usepackage{algorithm}%
\usepackage{algorithmicx}%
\usepackage{algpseudocode}%
\usepackage{listings}%
\usepackage[super]{nth}
\usepackage[version=4,arrows=pgf]{mhchem}%
\usepackage{siunitx}

\usepackage{physics}
\usepackage{lmodern}
\usepackage[utf8]{inputenc}
\usepackage{multirow}

\usepackage[normalem]{ulem}
\usepackage{graphicx}
\usepackage{ bbold }
\usepackage{hyperref}
\usepackage{caption}
\usepackage{subcaption}
\usepackage{physics}

\usepackage{xcolor}

\newcommand\chain{$[001]$}
\newcommand\pchain{$[ 1\Bar{1}0]$}

\usepackage{tabularx, makecell, array, booktabs}  % In the preamble

\newcolumntype{Y}{>{\centering\arraybackslash}X}
\begin{document}

%\title{Moiré-assisted charge instability in ultrathin \ce{RuO2}}
\title{{Moiré-Resonant Surface State in Ultrathin \ce{RuO2(110)}}}

%%=============================================================%%
%% GivenName	-> \fnm{Joergen W.}
%% Particle	-> \spfx{van der} -> surname prefix
%% FamilyName	-> \sur{Ploeg}
%% Suffix	-> \sfx{IV}
%% \author*[1,2]{\fnm{Joergen W.} \spfx{van der} \sur{Ploeg} 
%%  \sfx{IV}}\email{iauthor@gmail.com}
%%=============================================================%%

\author[1,2]{\fnm{Philipp} \sur{Keßler}}

\author[2,3]{\fnm{Andreas} \sur{Feuerpfeil}}

\author[2,3,4]{\fnm{Armando} \sur{Consiglio}}

\author[2,3]{\fnm{Hendrik} \sur{Hohmann}}

%\author[4]{\fnm{Domenico} \sur{Di Sante}}

\author[2,3]{\fnm{Ronny} \sur{Thomale}}

\author[1,2]{\fnm{Jonas} \sur{Erhardt}}

\author[1,2]{\fnm{Bing} \sur{Liu}}

\author[5,6]{\fnm{Vedran} \sur{Jovic}}

\author[1,2]{\fnm{Ralph} \sur{Claessen}}

\author[1,2]{\fnm{Patrick} \sur{Härtl}}

\author[2,3,7]{\fnm{Matteo} \sur{Dürrnagel}}

\author*[8,1,2]{\fnm{Simon} \sur{Moser}}\email{simon.moser@ruhr-uni-bochum.de}

\affil[1]{\orgdiv{Physikalisches Institut}, \orgname{Universität Würzburg}, \orgaddress{ \city{Würzburg}, \postcode{97074}, \country{Germany}}}

\affil[2]{\orgdiv{Würzburg-Dresden Cluster of Excellence ctd.qmat}, \orgaddress{ \city{Würzburg}, \postcode{97074}, \country{Germany}}}

\affil[3]{\orgdiv{Institut für Theoretische Physik und Astrophysik}, \orgname{Universität Würzburg}, \orgaddress{ \city{Würzburg}, \postcode{97074}, \country{Germany}}}

\affil[4]{\orgdiv{Istituto Officina dei Materiali}, \orgname{CNR-IOM}, \orgaddress{ \city{Trieste}, \postcode{34139}, \country{Italy}}}

\affil[5]{\orgdiv{Earth Resources and Materials}, \orgname{Institute of Geological and Nuclear Science}, \orgaddress{ \city{Lower Hutt}, \postcode{5010}, \country{New Zealand}}}

\affil[6]{\orgdiv{Boston University}, \orgname{Department of Physics}, \orgaddress{ \city{Boston}, \postcode{MA 02215}, \country{USA}}}

\affil[7]{\orgdiv{Institute for Theoretical Physics}, \orgname{ETH Zürich}, \orgaddress{ \city{Zürich}, \postcode{8093}, \country{Switzerland}}}

\affil[8]{\orgdiv{Experimentalphysik IV - AG Oberﬂächen}, \orgname{Ruhr-Universität Bochum}, \orgaddress{\city{44801 Bochum},  \country{Germany}}}

 \abstract{\ce{RuO2} has emerged as a prototypical candidate for \textit{altermagnetism}. In the face of daunting evidence for magnetic order in the bulk, the focus naturally shifted to surfaces and ultrathin films, where Coulomb interactions are dimensionally quenched and electron correlations strongly enhanced. Here, we examine atomically ordered, ultrathin \ce{RuO2(110)} grown on \ce{Ru(0001)} using a combination of scanning tunneling microscopy (STM), density functional theory, and density matrix renormalization group methods. We observe a nonmagnetic charge order that is imprinted by the incommensurate moiré stacking with the substrate and enhanced by the electronic Fermi surface scattering within the flat-band surface state. We further identify a nonmagnetic, metastable $c(2 \times 2)$ surface reconstruction that arises from surface phonon softening and can be toggled reversibly via STM tip manipulation. Spin-polarized STM measurements, however, reveal no evidence of magnetic order on the \ce{RuO2(110)} surface. Our findings of a nonmagnetic charge-modulation position ultrathin \ce{RuO2(110)} as an intriguing platform for exploring moiré-assisted electronic orders.}

%\keywords{\ce{RuO2}, Moiré, Flat surface state, Fermi surface instability, Altermagnetism, STM, ARPES, DMRG, DFT}

\maketitle

\begin{figure*}[t]
\centering
\includegraphics[width=1\textwidth]{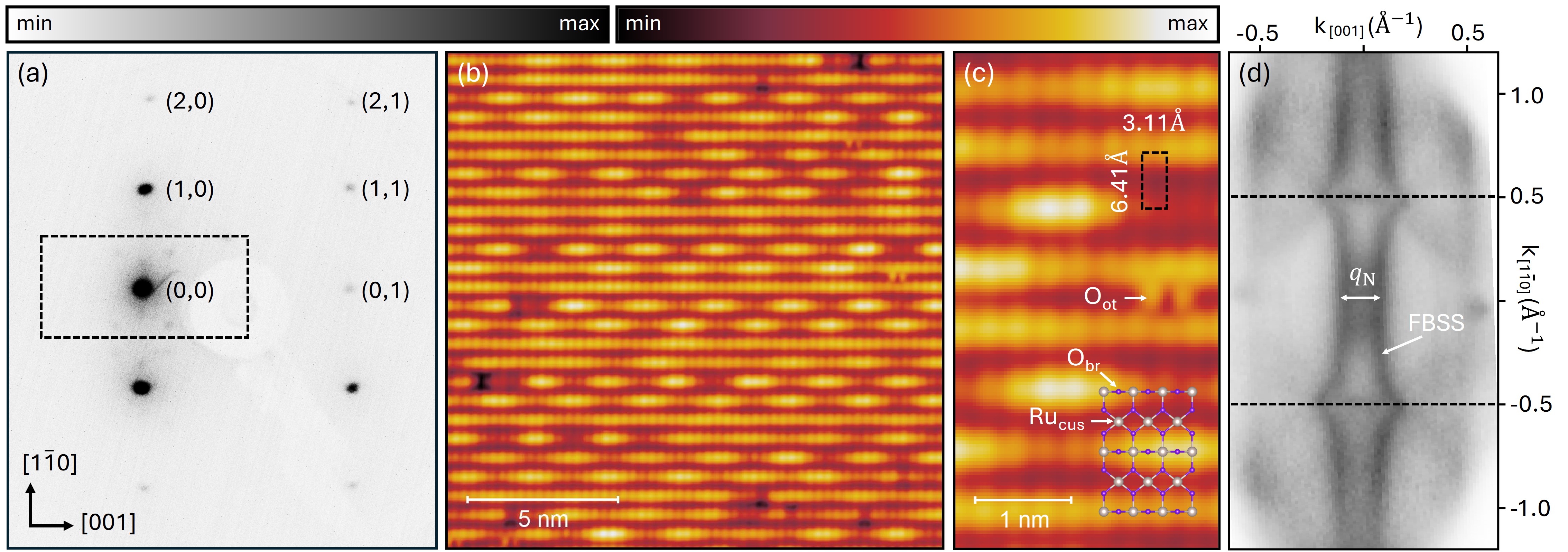}
\caption{Surface characterization of the stoichiometric \ce{RuO2}(110) thin film grown on \ce{Ru}(0001): (a) Room temperature LEED measurement ($E_\textrm{Kin} = \SI{80}{eV}$) of a large single domain of \ce{RuO2}(110); the surface Brillouin zone is indicated in black. (b) STM constant current topography map showing an extended \ce{RuO2(110)} terrace ($I_\text{T} = \SI{500}{pA}$, $V_\text{bias} = \SI{5}{mV}$). (c) Magnified view of panel (b) overlaid with the ball-and-stick model of the unreconstructed stoichiometric \ce{RuO2}(110), exposing \chain-oriented \ce{Ru-O_{br}} chains separated by rows of undercoordinated \ce{Ru_{cus}} \cite{Over2012}. Oxygen atoms are marked in purple and ruthenium atoms in gray. (d) ARPES Fermi surface map from a single domain, measured at the MAESTRO beamline of the Advanced Light Source with  $\SI{69}{eV}$ $p$-polarized photons at \SI{<3 E-10}{mbar} and \SI{40}{K}.}\label{fig:Char}
\end{figure*}

The binary oxide \ce{RuO2} has emerged as a material of multifold significance, drawing attention both for its exceptional catalytic properties and its proposed role as a functional Dirac semimetal with allegedly unconventional magnetic behavior~\cite{Over2012,Sun2017,Jovic2021,Feng2022}. Historically valued for its performance in electrochemical water splitting and its stability as a contact material~\cite{Over2012,Lee2012,Vadimsky1979}, \ce{RuO2} has more recently been identified as a potential prototype of \textit{altermagnetism}--a novel magnetic phase where the presence of alternating magnetic moments within the unit cell combined with the absence of an anti-unitary symmetry enables spin-split magnon excitations and thus hybridizes characteristics of conventional ferro- and antiferromagnets~\cite{Smejkal2022a,Smejkal2022b}.

Although several experimental studies have reported findings not inconsistent with an altermagnetic interpretation~\cite{Feng2022,Tschirner2023,Zhou2024,Guo2024,Jeong2025}, the existence of intrinsic magnetism in stoichiometric \ce{RuO2} remains highly contested. Theoretical work has suggested that strong Fermi surface nesting in \ce{RuO2} could drive spin-density wave instabilities~\cite{Mattheiss1976,Sun2017,Jovic2018,Jovic2019,Ahn2019}, and early neutron and resonant x-ray scattering studies appeared to support collinear magnetic order~\cite{Berlijn2017,Zhu2019,Lovesey2022}. Recent and more sensitive muon spin rotation and neutron scattering measurements, however, have excluded the presence of long-range magnetic order or sizable local moments in both bulk and thin film samples~\cite{Smolyanyuk2024,Kessler2024a,Kiefer2024}. Moreover, transport anomalies initially attributed to an anomalous Hall effect~\cite{Feng2022,Tschirner2023} can now be reconciled from nonmagnetic mechanisms~\cite{Pawula2024,Peng2024}, while purported time-reversal symmetry breaking in the band structure~\cite{Fedchenko2024} has been challenged by the absence of measurable spin splitting~\cite{Liu2024}. Numerous additional studies have raised further skepticism about altermagnetism in bulk \ce{RuO2}~\cite{Plouff2025,Lovesey2023,Hahn2025,Smolyanyuk2025}.

As a result, recent theoretical efforts have increasingly focused on ultrathin films and surface-specific magnetic phenomena, with particular attention to the thermodynamically stable \ce{RuO2}(110) surface. This is because the associated dimensionality reduction, combined with an expected breakdown of the Fermi liquid paradigm due to enhanced electron correlations~\cite{Giamarchi2003}, has been promoted as a possible mechanism for surface magnetic instabilities~\cite{Brahimi2024,Jeong2024,Torun2013,Ho2025}. 

In this context, spin-polarized scanning tunneling microscopy and spectroscopy (SP-STM/STS) have been proposed as powerful tools to probe emergent surface magnetism with atomic-scale resolution~\cite{Hu2025,Sukhachov2024,Ho2025}. However, such spin-sensitive, atomically resolved measurements have remained experimentally challenging. A key limitation has been the absence of high-quality, atomically ordered \ce{RuO2}(110) surfaces: Despite yielding high-quality bulk films, conventional epitaxial methods--including magnetron sputtering, molecular beam epitaxy (MBE), and pulsed laser deposition (PLD)--often fall short of achieving the defect-free surface order required for large-scale STM/STS~\cite{Kessler2024b}.
A well-established alternative involves the controlled oxidation and reactive segregation of \ce{RuO2} on single-crystalline \ce{Ru}(0001), producing ultrathin, stoichiometric \ce{RuO2(110)} films with excellent atomic order and minimal defect density~\cite{Over2000,Over2012}. This approach offers two key advantages: (i) it enables high-resolution STM/STS measurements that resolve the local electronic structure of the \ce{RuO2}(110) surface; and (ii) self-passivation restricts \ce{RuO2} to thicknesses where correlation effects are expected to be most pronounced, potentially giving rise to emergent electronic or spin instabilities.

Our results reveal surface-terminating chains of \ce{Ru} and \ce{O} that host a nested, quasi-one-dimensional flat-band surface state (FBSS), characterized by a local density of states (LDOS) peak near the Fermi energy--consistent with prior angle-resolved photoemission spectroscopy (ARPES) observations on bulk-truncated \ce{RuO2}(110)~\cite{Jovic2018,Jovic2019,Jovic2021,Visscher2026}. This FBSS is modulated by a moiré potential originating from the underlying ruthenium substrate, the strength of which can be tuned by varying the \ce{RuO2} film thickness.

The moiré potential induces a surface charge modulation that resonates with Fermi surface nesting of the FBSS, without evidence for a corresponding spin-modulation: within its limits, spin-polarized STM reveals no magnetic contrast, consistent with our Density Matrix Renormalization Group (DMRG) calculations. Additionally, we identify a metastable, previously unreported $c(2 \times 2)$ surface reconstruction that can be reversibly toggled using the STM tip and originates from a structural relaxation, as corroborated by density functional theory (DFT). Together, these findings reveal intriguing complex structural and electronic phenomena on the \ce{RuO2(110)} surface yet argue against its propensity towards magnetic instabilities.

\section*{Fermi Surface Nesting in \ce{RuO2}(110)}

The $C_{3v}$ symmetry of the \ce{Ru(0001)} substrate permits the growth of three rotational domains of \ce{RuO2}(110). Fine-tuning of the growth conditions~\cite[Secs.~S1,~S2]{supp}, however, achieves selective formation of millimeter-scale single-domain \ce{RuO2}(110) films. Low-energy electron diffraction (LEED, Fig.~\ref{fig:Char}(a)), confirms the long-range crystalline order, while atomically resolved STM topography images [Fig.~\ref{fig:Char}(b)] reveal well-ordered terraces with typical widths of several tens of nanometers.

A magnified view [Fig.~\ref{fig:Char}(c)] resolves the characteristic surface structure, consistent with the ball-and-stick model of the stoichiometric \ce{RuO2}(110) surface~\cite{Over2012}. In particular, the surface features quasi-one-dimensional \chain-oriented \ce{Ru-O_{br}} chains separated by rows of undercoordinated \ce{Ru_{cus}} sites. As previously reported~\cite{Jovic2018,Ho2025,Jovic2021}, the hybridization of \ce{Ru} $4d_{z}$ and \ce{O_{br}} $2p_{z}$ orbitals in these chains gives rise to a FBSS that disperses along the \chain-direction and is flat in \pchain-direction. This FBSS is clearly observed in ARPES measurements on our ultrathin, $\lesssim\SI{3}{nm}$ \ce{RuO2}(110)/\ce{Ru(0001)} films [Fig.~\ref{fig:Char}(d)]. Notably, ARPES displays the canonical (110)-oriented Fermi surface known from cleaved bulk \ce{RuO2}~\cite{Jovic2018,Liu2024,Visscher2026}, indicating that the band structure is largely unaffected by quantum confinement at this thickness. The Fermi surface is dominated by two linear features of the FBSS oriented along the \pchain\ direction. Its near-parallel geometry enables strong Fermi surface nesting, suggesting an enhanced susceptibility to electronic instabilities and a possible breakdown of Fermi liquid behavior due to electron interactions~\cite[Sec.~S6.2]{supp}. The corresponding nesting vector is determined to be $q_\textrm{N} \approx \SI{0.24 \pm 0.04}{\AA^{-1}}$.

\begin{figure*}
\centering
\includegraphics[width=1\textwidth]{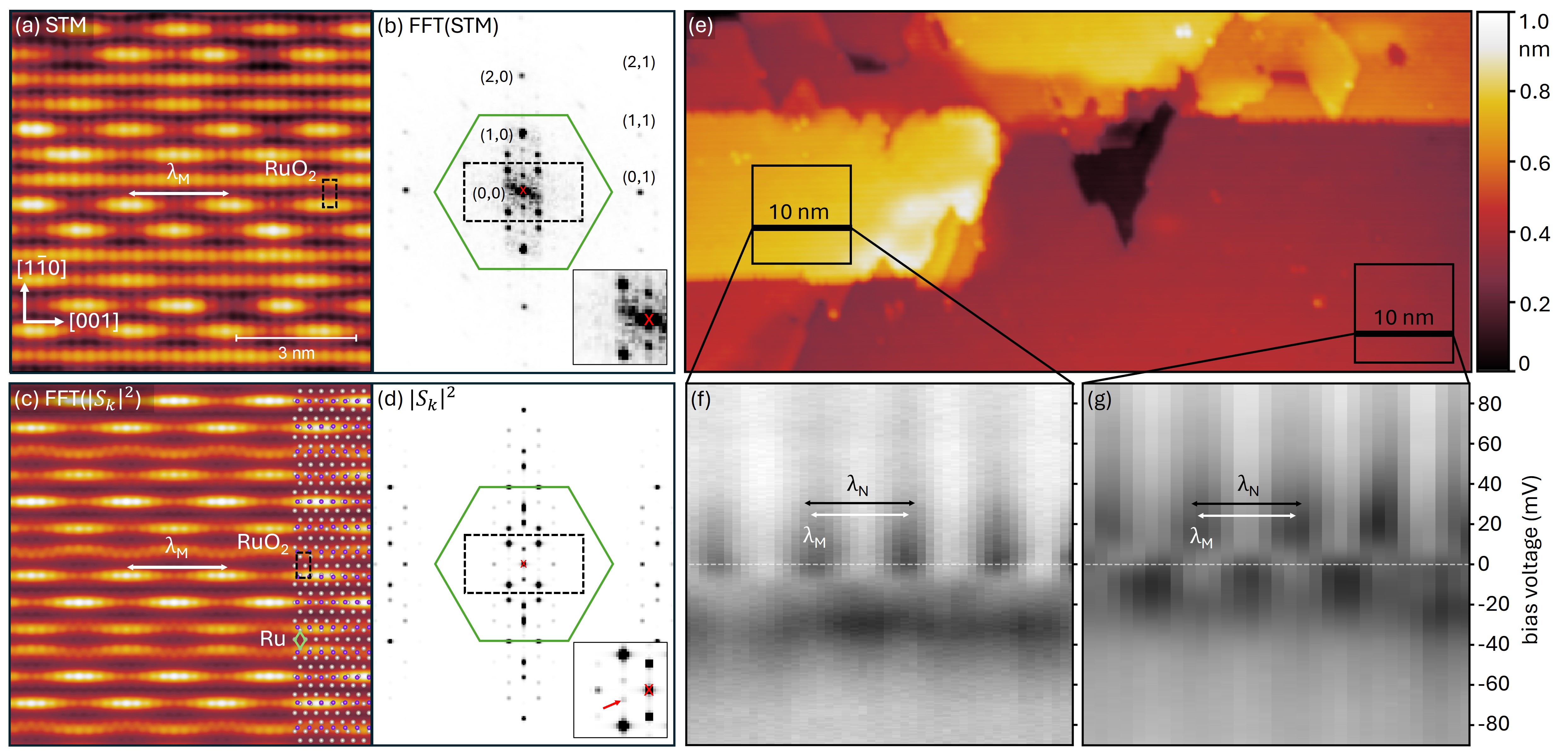}
\caption{Moiré pattern arising from the lattice mismatch of the \ce{RuO2(110)} thin film and the \ce{Ru(0001)} substrate: (a) STM constant current topography map showing real-space moiré pattern ($I_\text{T} = \SI{500}{pA}$, $V_\text{bias} = \SI{5}{mV}$). (b) FFT of the STM measurement depicted in panel (a). The green and black lines mark the boundaries of the \ce{Ru(0001)} and \ce{RuO2(110)} surface Brillouin zones, respectively. (c) Real-space simulation of the moiré pattern, i.e., FFT of the structure factor $|S_{\vb{q}}|^2$ in (d) calculated via Eq.~\eqref{eqn:structure_factor}. The Ru atom positions of the \ce{Ru(0001)} substrate and \ce{RuO2(110)} surface are indicated by grey and purple circles respectively. (e) STM constant current topography map exposing different apparent heights of the \ce{RuO2} film ($I_\text{T} = \SI{10}{pA}$, $V_\text{bias} = \SI{-500}{mV}$). Corresponding d$I$/d$V$ STS line spectroscopy along the \chain-direction of the two different regions is shown in panels (f) and (g) ($I_\text{T} = \SI{10}{pA}$, $V_\text{bias} = \SI{-500}{mV}$, $V_\text{rms} = \SI{2}{mV}$, $\delta_z = \SI{-0.27}{nm}$).}
\label{fig:Moire}
\end{figure*}

\section*{Substrate-Induced Moiré Modulation}

Having established the synthesis of atomically ordered \ce{RuO2(110)} surfaces with an electronic structure consistent with cleaved single crystals, we now turn to two notable features observed in the STM topography of Fig.~\ref{fig:Char}(b,c)~\cite{Over2000,Over2002a,Rossler2007a}:\footnote{According to a private communication with Michael Schmid (TU Vienna), the authors of Ref.~\cite{Over2000} occasionally observed a periodic corrugation in ultrathin \ce{RuO2(110)}/Ru(0001) films using room-temperature STM at a bias voltage of $\SI{-30}{mV}$. This corrugation is consistent with our observations (ii), although these results have not been published.}

\begin{enumerate}
    \item[(i)] The registry shift between some occasional on top \ce{O_{ot}} atoms and the \ce{Ru-O_{br}} chains~\cite{Over2002a} indicates that our low bias voltage ($V_{\text{bias}} = \SI{5}{mV}$) STM measurements predominantly resolve the \ce{Ru}-positions. This contrasts earlier work at higher bias voltages ($\sim\SI{-1}{V}$) that resolved the \ce{O_{br}} positions~\cite{Over2000}.

    \item[(ii)] The \ce{Ru-O_{br}} chains, which host the FBSS, show a distinct apparent height beating in the order of $\SI{30}{pm}$. The magnitude and phase of this beating varies from chain to chain.    
\end{enumerate}

While the sensitivity of low bias voltages to \ce{Ru}-derived density of states~\cite{Jovic2021} explains point (i), point (ii) requires a more detailed examination: Applying a 2D fast Fourier transform (FFT) to the quasi-periodic modulation observed in the STM topography [Fig.\ref{fig:Moire}(a)] reveals additional peaks within the first Brillouin zone [Fig.~\ref{fig:Moire}(b)], that are inconsistent with the \ce{RuO2(110)} lattice alone. These are rationalized by the incommensurate lattice of the underlying \ce{Ru(0001)} substrate, which imprints a moiré potential on the \ce{RuO2(110)} surface~\cite{Kim2000,Kim2001a} [see inset of Fig.~\ref{fig:Moire}(c)]. This induces a surface charge density modulation that can be expressed in terms of a moiré structure factor~\cite[Sec.~S3]{supp}:
\begin{equation}
S_{\vb{q}} = \sum_{\vb{G}_{\ce{Ru}}} \sum_{\vb{G}_{\ce{RuO2}}} \delta(\vb{q} - \vb{G}_{\ce{Ru}} - \vb{G}_{\ce{RuO2}}) f(G_{\ce{RuO2}})\;,
\label{eqn:structure_factor}
\end{equation}
with atomic scattering factor $f(G_{\ce{RuO2}})$ \cite{Peng1999}, and reciprocal lattice vectors $\vb{G}_{\ce{RuO2}}$ and $\vb{G}_{\ce{Ru}}$ of film and substrate.

The calculated $|S_{\vb{q}}|^2$ [Fig.~\ref{fig:Moire}(d)] matches remarkably well with the experimental FFT [Fig.~\ref{fig:Moire}(b)], as well as the LEED pattern [$E_\textrm{Kin} = \SI{80}{eV}$, $T=\SI{300}{K}$, Fig.~\ref{fig:Char}(a)]. To compare with real-space data, we performed an inverse FFT of Fig.~\ref{fig:Moire}(d), applying a Gaussian filter to simulate STM resolution. The resulting image [Fig.~\ref{fig:Moire}(c)] is in excellent agreement with the measured topography [Fig.~\ref{fig:Moire}(a)], revealing a leading moiré wavelength of $\lambda_\textrm{M} = \SI{22.9\pm 2.0}{\AA}$, i.e., $ \approx 7.6$ lattice constants along the chain direction.

To investigate the thickness dependence of the moiré modulation, we prepared a stepped \ce{RuO2(110)} film with terraces of varying apparent heights. A representative STM image [Fig.~\ref{fig:Moire}(e)] shows an apparently thicker region (bright orange) and an apparently thinner region (dark orange).

In the thicker region, spatially resolved d$I$/d$V$ spectra [Fig.~\ref{fig:Moire}(f)] reveal a modulated LDOS, with peaks near zero bias. They alternate in intensity with the spatial period of $\lambda_\textrm{M}$ --- a clear fingerprint of moiré induced LDOS trapping. Additionally, a second LDOS feature appears at \SI{-30}{meV}, which shows almost no spatial modulation and aligns with the FBSS binding energy observed in ARPES~\cite[Figs.~S1,~S2(c)]{supp}. This indicates a substrate-decoupled surface electronic structure of \ce{RuO2(110)} in the thicker region \cite{Jovic2018}.

In contrast, the thinner region shows these LDOS features shifted upward by \SI{\approx20}{meV}, now located at bias voltages of \SI{-10}{meV} and \SI{20}{meV} [Fig.~\ref{fig:Moire}(g)]. Both signatures exhibit a clear, inversely phased spatial modulation with period $\lambda_\textrm{M}$, indicating stronger coupling to the underlying \ce{Ru}(0001) substrate and a more pronounced influence of the moiré potential. This suggests a progressive attenuation of the moiré potential with increasing film thickness and underscores the decisive role of the substrate in shaping the \ce{RuO2(110)} thin film's surface electronic structure.

\begin{figure*}
\centering
\includegraphics[width=1\textwidth]{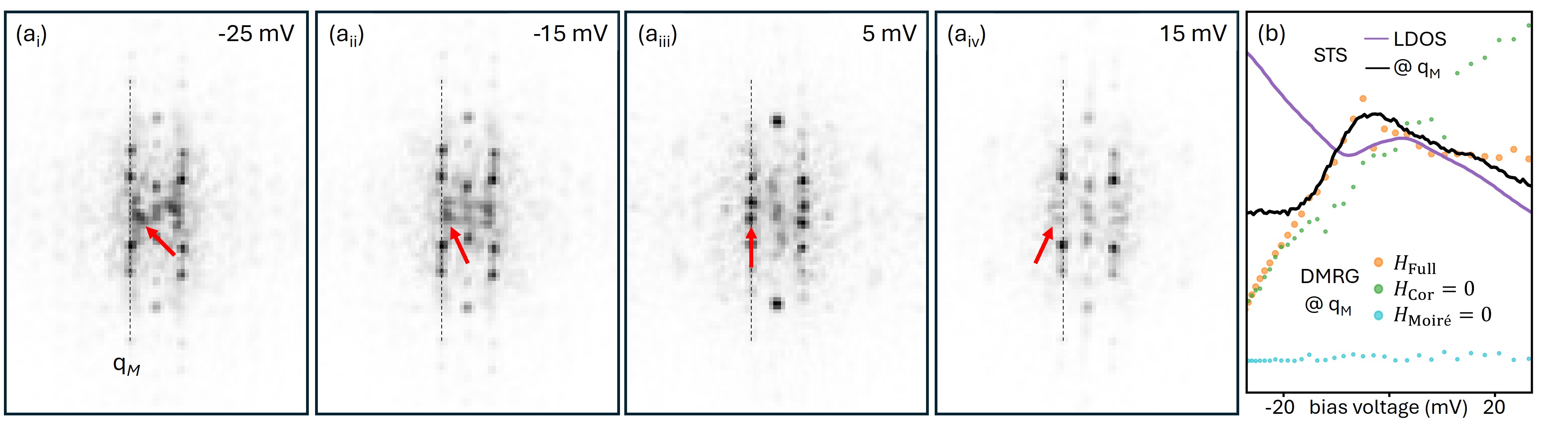}
\caption{Moiré interaction with FBSS: (a) FFT of constant current d$I$/d$V$ maps collected at bias voltages from $V_\text{bias} = \SI{-25}{mV}$ to $\SI{15}{mV}$ ($I_\text{T} = \SI{500}{pA}$, $V_\text{rms} = \SI{5}{mV}$). The red arrow indicates a dispersing moiré peak. (b) LDOS integrated over the entire area (purple) of an STS grid measurement, as compared to LDOS integrated only over the moiré frequencies (black) at $q_\textrm{M}$ as indicated in panel (a), (the grid scan was measured on the thicker region of Fig. \ref{fig:Moire}(e) with $I_\text{T} = \SI{10}{pA}$, $V_\text{bias} = \SI{-500}{mV}$, $V_\text{rms} = \SI{2}{mV}$, $\delta_z = \SI{-0.27}{nm}$). While the total LDOS exhibits maxima at $\SI{-30}{mV}$ and near zero bias, the moiré features exhibit a resonance at $\SI{-5}{mV}$. The markers show the oscillation amplitude at $q_\textrm{M}$ calculated by DMRG for the FBSS without moiré (turquoise), for non-interacting FBSS and moiré (green) and for interacting FBSS and moiré (orange), the latter of which reproduces the experimentally observed resonance (black).}
\label{fig:Spectro}
\end{figure*}

\section*{Resonating Fermi surface scattering}

Building on our interpretation of the moiré-induced surface charge modulation, we note that a contrast inversion between STM images taken at opposite bias polarities---similar to Fig.~\ref{fig:Moire}(g)---is often regarded as a hallmark of charge density wave order~\cite{Rodriguez1999,Mallet2001,Spera2020}, i.e., of a Fermi surface instability. We, however, reconcile this modulation by a resonant coupling between the moiré potential and the Fermi-surface nesting, corroborated by a striking correspondence: the moiré wave vector $q_\textrm{M} = 2\pi/\lambda_\textrm{M} \approx \SI{0.27\pm0.02}{\AA^{-1}}$, as extracted from STM data, closely matches the Fermi surface nesting vector $q_\textrm{N} \approx \SI{0.24\pm0.04}{\AA^{-1}}$ identified within the FBSS via ARPES [Fig.~\ref{fig:Char}(d)]~\cite{Jovic2018}. This near-coincidence naturally implies a contribution of the Fermi surface nesting within the FBSS to the observed moiré scattering features:

Indeed, we find that the electronic fluctuations encoded in the bare susceptibility $\chi(\mathbf{q})$ can constructively interfere with the moiré potential when both wave vectors are commensurate. In this sense, the Coulomb interaction can drive a resonant response by coupling the moiré potential to $\chi(\mathbf{q})$, which itself is a property of the non-interacting band structure.

To explore this possibility experimentally, Fig.~\ref{fig:Spectro}(a) presents two-dimensional Fourier transforms (FFTs) of d$I$/d$V$ maps, which reflect the local density of states near the Fermi level, acquired at bias voltages of $V_{\text{bias}} = \SI{-25}{mV}$, $\SI{-15}{mV}$, $\SI{5}{mV}$, and $\SI{15}{mV}$. As expected, the FFTs qualitatively reproduce the peak pattern associated with the static moiré modulation observed in the STM topography [Fig.~\ref{fig:Moire}(c)]. Notably, however, the intensity profile of the corresponding moiré scattering peaks at $q_M$ [black curve in Fig.~\ref{fig:Spectro}(b)] -- measured by STS-grid spectroscopy -- exhibit a pronounced bias-voltage dependence, reaching a maximum at $-5\,\mathrm{meV}$. This behavior is qualitatively distinct from a corresponding dip in the total surface density of states, which we obtained by integration over all wavevectors of the dataset (purple curve). This discrepancy indicates that the resonance is not driven by the LDOS itself, but instead reflects a moiré-induced interaction with Fermi surface nesting in the FBSS, occurring at band fillings where $q_\textrm{M}$ and $q_\textrm{N}$ coincide--a phenomenon that, to the best of our knowledge, has not been previously reported.

\begin{figure}
\centering
\includegraphics[width=1\columnwidth]{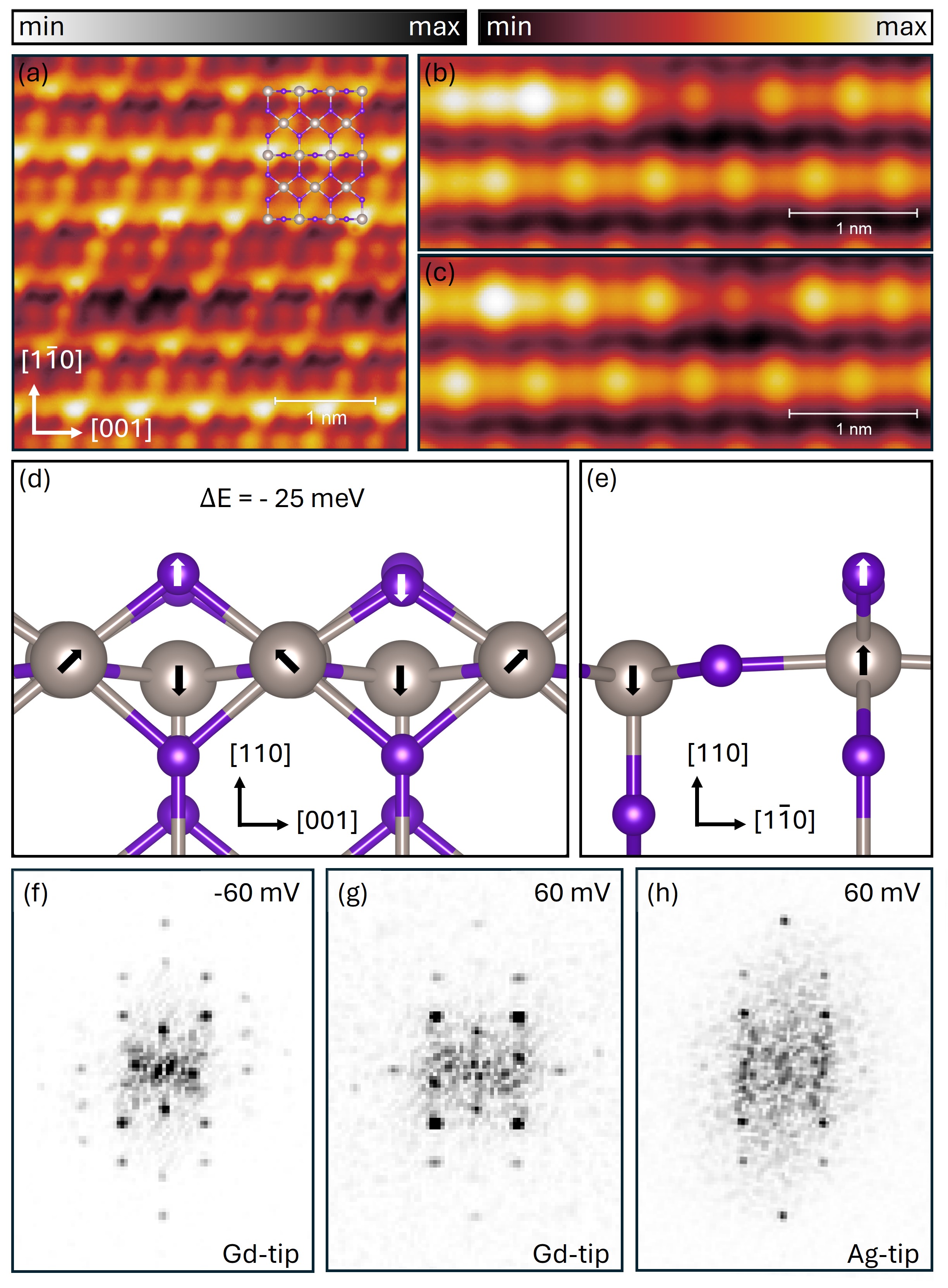}
\caption{\ce{RuO2(110)} surface instabilities: (a) Constant current STM topography scan showing a $c(2 \times 2)$ reconstruction of the \ce{RuO2(110)} surface ($I_\text{T} = \SI{1000}{pA}$, $V_\text{bias} = \SI{-15}{mV}$). (b,c) Constant current STM topography scans on the same area taken before and after switching between the two metastable reconstructions ($I_\text{T} = \SI{20}{pA}$, $V_\text{bias} = \SI{-15}{mV}$). (d,e) $c(2 \times 2)$ surface reconstruction of \ce{RuO2(110)} as found by DFT slab calculations. (f,g) FFT of constant current d$I$/d$V$ scans measured with spin polarized Gd-tip at (f) $V_\text{bias} = \SI{-60}{mV}$ and (g) $V_\text{bias} = \SI{60}{mV}$ ($I_\text{T} = \SI{200}{pA}$, $V_\text{rms} = \SI{5}{mV}$). (h) Corresponding FFT of unpolarized Ag-tip scan showing qualitatively similar results, indicating the peak's origin to be non-magnetic.}\label{fig:Reconstructions}
\end{figure}

To theoretically examine this resonance, we employed the DMRG method, well-suited for probing quasi-one-dimensional strongly correlated systems~\cite{White_1992,SCHOLLWOCK201196}. We constructed a minimal Hamiltonian model incorporating key features of the \ce{RuO2 (110)} surface: (i) a tight-binding term ($H_\text{Kin}$) derived from the DFT surface band structure and adapted to the FBSS ARPES results~\cite{Jovic2018,Jovic2021}, (ii) a periodically modulated onsite energy representing the moiré potential ($H_\text{Moiré}$), and (iii) a two-particle interaction term capturing electronic correlations ($H_\text{Cor}$). 
The total Hamiltonian reads~\cite[Sec.~S6.1]{supp}:

\begin{equation}
   H_\text{Full} = H_\text{Kin} + H_\text{Moiré} + H_\text{Cor}. 
\end{equation}

The prominent anisotropy of the electronic single-particle spectrum in Fig.~\ref{fig:Char}(d) suggests weakly coupled \ce{Ru-O_{br}} chains to dominate the low energy physics of the \ce{RuO2}(110) surface. This intuition is substantiated by the almost one dimensional electronic fluctuation spectrum characterized by the bare susceptibility $\chi(\mathbf q)$~\cite[Sec.~S6.2]{supp} and motivates us to start our analysis by considering an isolated \ce{Ru-O_{br}} chain along the \chain-axis. Using open boundary conditions, we solve the model for varying FBSS binding energies by tuning the chemical potential. As an analogue to the experimental LDOS, we compute the expectation value of the site-resolved electronic density $\langle \hat{n}_i \rangle = \langle \hat{n}_{i\uparrow} + \hat{n}_{i\downarrow} \rangle$.

Consistent with experiment, $\langle \hat{n}_i \rangle$ displays a periodic modulation along the investigated 1D chain~\cite[Fig.~S9(c)]{supp}. Its FFT amplitude, plotted as a function of energy and momentum, reveals two main features: a dominant, non-dispersive peak at $q_\textrm{M} = 2\pi/\lambda_\textrm{M}=\SI{0.27}{\AA^{-1}}$, matching the experimentally determined moiré period, and a quadratically dispersing mode associated with Friedel oscillations from the FBSS~\cite[Sec.~S6.4]{supp}\cite{White2002}. The strength of the charge modulation at $q_\textrm{M}$ is plotted as function of binding energy in Fig.~\ref{fig:Spectro}(b) (orange curve), and indeed reproduces the characteristic resonance at $\SI{-5}{mV}$ observed experimentally (black curve), where the Fermi wavevector $q_\textrm{N}$ perfectly matches the moiré period $q_\textrm{M}$.

While in the real material the moiré potential and Fermi-surface nesting are intrinsically intertwined, the DMRG framework allows us to disentangle their respective roles~\cite[Sec.~S6.4]{supp}. When Fermi-surface scattering is suppressed by setting $H_{\text{Cor}} = 0$, a charge modulation with the correct periodicity $q_{\mathrm{M}}$ is recovered; however, the resonant enhancement is absent (green curve). Conversely, when the moiré potential is switched off ($H_{\text{Moiré}} = 0$), electronic correlations alone are insufficient to drive a charge instability at $q_{\mathrm{M}}$, and only residual Friedel oscillations arising from the open boundary conditions remain (turquoise curve).
\\
Extending our 1D analysis to the full 2D surface model, we find that charge ordering along the \chain-axis is indeed consistent with the resonant behavior of the isolated chains. The inter-chain ordering is governed exclusively by the moiré potential $H_{\text{Moiré}}$, whereas the weak inter-chain coupling contained in $H_{\text{Kin}}$ and $H_{\text{Cor}}$ plays only a marginal role due to the pronounced anisotropy of the FBSS.
Taken together, these results demonstrate that the charge modulation is primarily driven by the moiré potential, while electronic correlations at the Fermi surface substantially enhance this modulation and give rise to the experimentally observed resonance in STS. Consequently, we do not expect charge ordering at $q_{\mathrm{M}}$ on the bulk-truncated \ce{RuO2}(110) surface.
\\
For completeness, let us also focus on the four distinct peaks that are indicated by red arrows in the FFTs of Fig.~\ref{fig:Spectro}(a), peaks that are predicted by our structure factor model [inset of Fig.~\ref{fig:Moire}(d)]. Unlike the other moiré peaks, these features exhibit a pronounced energy-dependent momentum shift: as the bias voltage increases from $\SI{-25}{mV}$ to $\SI{15}{mV}$, the peaks shift from $0.87q_\textrm{M}$ to $1.08q_\textrm{M}$, a currently unexplained energy dispersion that is about twice as steep as expected for quasi particle interference of the FBSS.

\section*{Emergent Surface Reconstruction}

In addition to the charge-ordering phenomena discussed above, STM topography measurements on a subset of samples [Fig.~\ref{fig:Reconstructions}(a)] reveal a pronounced $c(2\times2)$ periodic modulation. This reconstruction is characterized by an alternating intensity pattern of surface \ce{Ru} sites, effectively doubling the \ce{RuO2}(110) surface unit cell. 

As shown in Figs.~\ref{fig:Reconstructions}(b,c), two symmetry-equivalent metastable configurations of this $c(2\times2)$ pattern can be reversibly switched with tunneling currents of $I_\text{T} = \SI{100}{pA}$, demonstrating the bistable nature of the surface reconstruction. 
\\
A previous study under ambient conditions attributed a similar $2\times2$ reconstruction on \ce{RuO2}(110) to the adsorption of \ce{CO} molecules at oxygen bridge sites~\cite{Rossler2007a}. This interpretation, however, is incompatible with our observations, since \ce{CO} adsorption is known to quench the FBSS~\cite{Jovic2021}, which remains clearly discernible in our measurements. Instead, our DFT calculations for a clean \ce{RuO2}(110) surface slab reveal a structural relaxation that intrinsically doubles the surface unit cell [Figs.~\ref{fig:Reconstructions}(d,e)], is driven by a surface phonon softening, and \SI{\approx25}{meV} more stable than the unreconstructed $2\times2$ surface unit cell~\cite[Sec.~S6.5]{supp}.

\section*{No Evidence for Magnetic Surface Instability}

Finally, none of our experimental observations provide evidence for magnetic order in the \ce{RuO2} thin films, consistent with the non-magnetic ground states obtained from DMRG. Even when considering on-site Coulomb repulsions exceeding both FBSS bandwidth and \emph{ab initio} bulk values~\cite{Smolyanyuk2024}, neither of the considered models exhibits any tendency towards long-range magnetic order in the spin expectation value $S_i^z = \langle \hat{n}_{i\uparrow} - \hat{n}_{i\downarrow} \rangle / 2$. Instead, all computed ground states remain spin singlets without any local magnetization.
\\
In light of recent theoretical predictions of surface magnetism on \ce{RuO2}(110)~\cite{Torun2013,Jeong2024,Ho2025}, we explicitly probed potential magnetic order using spin-polarized STM measurements performed with a \ce{Gd}-coated tip. The magnetic sensitivity of the tip was independently verified on a \ce{Gd} reference surface~\cite[Fig.~S6]{supp}. Within the experimental resolution, no significant differences were observed between data acquired with the spin-polarized tip [Figs.~\ref{fig:Reconstructions}(f,g)] and those obtained using an unpolarized Ag tip [Fig.~\ref{fig:Reconstructions}(h)]. In particular, we find no indications of spin-density waves, magnetic domains, or spin-polarized surface states.
\\
We note that the sub-atomic magnetic order with small magnetic moments predicted in the context of altermagnetism in Ref.~\cite{Ho2025} approaches the sensitivity limits of spin-polarized STM (spatial resolution $\approx \SI{3}{\angstrom}$, magnetic resolution $\approx \SI{0.3}{\mu_{\mathrm{B}}}$)~\cite{Tersoff1983,Kurz2002,hartl2022,Chen1990,Chen1990b}. The absence of a measurable spin contrast therefore constitutes an important experimental benchmark for future studies addressing possible magnetic ordering on \ce{RuO2}(110).
\\
Experimental signatures of altermagnetism in epitaxial \ce{RuO2} thin films have primarily been reported for \ce{RuO2} grown on \ce{TiO2}(110), where substantial strain and symmetry breaking are imposed on the (110) surface~\cite{Rajapitamahuni2024,Jeong2024,Jeong2025,Zhang2025,Akashdeep2025}. By contrast, Ref.~\cite{Hahn2025} reports ARPES measurements on \ce{RuO2} thin films grown on metallic Ru substrates, which, similar to our results, are consistent with a non-magnetic surface band structure. This apparent substrate dependence raises questions about the intrinsic nature of altermagnetism in \ce{RuO2} thin films and highlights the need for a more systematic investigation of substrate-induced effects.
\\
Irrespective of this open issue, our study establishes ultrathin \ce{RuO2} films as a versatile platform for correlated electron physics, in which tunable electronic and structural instabilities intertwine to give rise to ordering phenomena and surface reconstructions that, within the limits of our experimental resolution, remain non-magnetic.

\section*{Acknowledgements}
We thank T. Müller, L. Klebl and M. Bode for valuable discussions. Funding support came from the Deutsche Forschungsgemeinschaft (DFG, German Research Foundation) under Germany’s Excellence Strategy through the Würzburg-Dresden Cluster of Excellence for Complexity, Topology and Dynamics in Quantum Matter ctd.qmat (EXC 2147, Project ID 390858490) and through the Collaborative Research Center SFB 1170 ToCoTronics (Project ID 258499086), as well as from the New Zealand Ministry of Business, Innovation and Employment (MBIE, Grant number: C05X2004). A.F. thanks the CCQ at the flatiron institute for their hospitality. M.D. is grateful for support from a Ph.D. scholarship of the Studienstiftung des deutschen Volkes. A.C. acknowledges support from PNRR MUR project PE0000023-NQSTI. A.F. and A.C. further acknowledge the Gauss Centre for Supercomputing e.V. (https://www.gauss-centre.eu) for funding this project by providing computing time on  the GCS Supercomputer SuperMUC-NG at Leibniz Supercomputing Centre (https://www.lrz.de). This research used resources of the Advanced Light Source, a U.S. DOE Office of Science User Facility under contract no. DE-AC02-05CH11231.\nocite{hartl2022,Schafer1963,Over2000,Over2004a,Over2012,Herd2012,Over2016,He2015,Jovic2018, Jovic2021,Kim2000,Kim2001,Kim2001a,Cotton1966, Swanson1955,Peng1999,Kresse1996,Kresse1999,Bloechl1994,Perdew1992,Perdew1996,White_1992,SCHOLLWOCK201196,White_2005,ITensor,ITensor-r0.3,Minkin2023,Brahimi2024,Jeong2024,Torun2013,Ho2025,Durrnagel2022,Chepiga2021,White2002,Basak2024,Choi2025,supp}

\bibliography{sn-bibliography}

@misc{supp,
  note = "See Supplemental Material at
    URL-will-be-inserted-by-publisher for details, which includes Refs.~[1, 3, 14, 32–35, 39, 41, 44–46, 50–52, 55, 61–81]."
}

@article{Tersoff1983,
   author = {J. Tersoff and D. R. Hamann},
   doi = {10.1103/PhysRevLett.50.1998},
   issn = {0031-9007},
   issue = {25},
   journal = {Physical Review Letters},
   month = {6},
   pages = {1998-2001},
   title = {{Theory and Application for the Scanning Tunneling Microscope}},
   volume = {50},
   url = {https://link.aps.org/doi/10.1103/PhysRevLett.50.1998},
   year = {1983}
}

@article{Chen1990,
   author = {C. Julian Chen},
   doi = {10.1103/PhysRevLett.65.448},
   issn = {0031-9007},
   issue = {4},
   journal = {Physical Review Letters},
   month = {7},
   pages = {448-451},
   title = {{Origin of atomic resolution on metal surfaces in scanning tunneling microscopy}},
   volume = {65},
   url = {https://link.aps.org/doi/10.1103/PhysRevLett.65.448},
   year = {1990}
}

@article{Chen1990b,
   author = {C. Julian Chen},
   doi = {10.1103/PhysRevB.42.8841},
   issn = {0163-1829},
   issue = {14},
   journal = {Physical Review B},
   month = {11},
   pages = {8841-8857},
   title = {{Tunneling matrix elements in three-dimensional space: The derivative rule and the sum rule}},
   volume = {42},
   url = {https://link.aps.org/doi/10.1103/PhysRevB.42.8841},
   year = {1990}
}

@article{Hahn2025,
   author = {Sungsoo Hahn and Minkyu Park and Yeonghoon Lee and Changsoo Kim and Jahyun Koo and Minhyuk Choi and Seungwoo Song and S. H. Rhim and Changyoung Kim and Chanyong Hwang},
   doi = {10.1103/zhm5-k3wr},
   issn = {2469-9950},
   issue = {21},
   journal = {Physical Review B},
   month = {6},
   pages = {214431},
   publisher = {American Physical Society},
   title = {{Off-stoichiometric surface reconstruction in the altermagnetic candidate \ce{RuO2(100)}/\ce{Ru(10$\bar{1}$ 0)}}},
   volume = {111},
   url = {https://link.aps.org/doi/10.1103/zhm5-k3wr},
   year = {2025}
}

@article{Kurz2002,
   author = {Ph Kurz and G Bihlmayer and S Blügel},
   doi = {10.1088/0953-8984/14/25/305},
   issn = {09538984},
   issue = {25},
   journal = {Journal of Physics: Condensed Matter},
   month = {7},
   pages = {305},
   title = {{Magnetism and electronic structure of hcp \ce{Gd} and the \ce{Gd(0001)} surface}},
   volume = {14},
   url = {https://iopscience.iop.org/article/10.1088/0953-8984/14/25/305},
   year = {2002}
}

@article{Jeong2025,
   author = {Seung Gyo Jeong and Seungjun Lee and Bonnie Lin and Zhifei Yang and In Hyeok Choi and Jin Young Oh and Sehwan Song and Seung wook Lee and Sreejith Nair and Rashmi Choudhary and Juhi Parikh and Sungkyun Park and Woo Seok Choi and Jong Seok Lee and James M. LeBeau and Tony Low and Bharat Jalan},
   doi = {10.1073/pnas.2500831122},
   issn = {0027-8424},
   issue = {24},
   journal = {Proceedings of the National Academy of Sciences},
   month = {6},
   pages = {e2500831122},
   title = {{Metallicity and anomalous Hall effect in epitaxially strained, atomically thin \ce{RuO2} films}},
   volume = {122},
   url = {https://pnas.org/doi/10.1073/pnas.2500831122},
   year = {2025}
}

@book{Giamarchi2003,
    author = {Giamarchi, Thierry},
    title = {{Quantum Physics in One Dimension}},
    publisher = {Oxford University Press},
    year = {2003},
    month = {12},
    isbn = {9780198525004},
    doi = {10.1093/acprof:oso/9780198525004.001.0001},
    url = {https://doi.org/10.1093/acprof:oso/9780198525004.001.0001},
}

@article{Mattheiss1976,
   author = {L. F. Mattheiss},
   doi = {10.1103/PhysRevB.13.2433},
   issn = {0556-2805},
   issue = {6},
   journal = {Physical Review B},
   month = {3},
   pages = {2433-2450},
   title = {{Electronic structure of \ce{RuO2}, \ce{OsO2}, and \ce{IrO2}}},
   volume = {13},
   url = {https://link.aps.org/doi/10.1103/PhysRevB.13.2433},
   year = {1976}
}

@article{hartl2022,
   author = {Patrick Härtl and Markus Leisegang and Matthias Bode},
   doi = {10.1103/PhysRevB.105.174431},
   issn = {2469-9950},
   issue = {17},
   journal = {Physical Review B},
   month = {5},
   pages = {174431},
   publisher = {American Physical Society},
   title = {{Magnetic domain structure of epitaxial \ce{Gd} films grown on \ce{W(110)}}},
   volume = {105},
   url = {https://link.aps.org/doi/10.1103/PhysRevB.105.174431},
   year = {2022}
}

@article{Perdew1996,
   author = {John P. Perdew and Kieron Burke and Matthias Ernzerhof},
   doi = {10.1103/PhysRevLett.77.3865},
   issn = {0031-9007},
   issue = {18},
   journal = {Physical Review Letters},
   month = {10},
   pages = {3865-3868},
   title = {{Generalized Gradient Approximation Made Simple}},
   volume = {77},
   url = {https://link.aps.org/doi/10.1103/PhysRevLett.77.3865},
   year = {1996}
}

@article{Kresse1996,
   author = {G. Kresse and J. Furthmüller},
   doi = {10.1103/PhysRevB.54.11169},
   issn = {0163-1829},
   issue = {16},
   journal = {Physical Review B},
   month = {10},
   pages = {11169-11186},
   title = {{Efficient iterative schemes for ab initio total-energy calculations using a plane-wave basis set}},
   volume = {54},
   url = {https://link.aps.org/doi/10.1103/PhysRevB.54.11169},
   year = {1996}
}

@article{Kresse1999,
   author = {G. Kresse and D. Joubert},
   doi = {10.1103/PhysRevB.59.1758},
   issn = {0163-1829},
   issue = {3},
   journal = {Physical Review B},
   month = {1},
   pages = {1758-1775},
   title = {{From ultrasoft pseudopotentials to the projector augmented-wave method}},
   volume = {59},
   url = {https://link.aps.org/doi/10.1103/PhysRevB.59.1758},
   year = {1999}
}

@article{Peng1999,
   author = {L.-M. Peng},
   doi = {10.1016/S0968-4328(99)00033-5},
   issn = {09684328},
   issue = {6},
   journal = {Micron},
   month = {12},
   pages = {625-648},
   title = {{Electron atomic scattering factors and scattering potentials of crystals}},
   volume = {30},
   url = {https://linkinghub.elsevier.com/retrieve/pii/S0968432899000335},
   year = {1999}
}

@article{Kessler2024b,
   author = {P. Keßler and T. Waldsauer and V. Jovic and M. Kamp and M. Schmitt and M. Sing and R. Claessen and S. Moser},
   doi = {10.1063/5.0217312},
   issn = {2166-532X},
   issue = {10},
   journal = {APL Materials},
   month = {10},
   pages = {101110},
   title = {{Epitaxial \ce{RuO2} and \ce{IrO2} films by pulsed laser deposition on \ce{TiO2}(110)}},
   volume = {12},
   url = {https://pubs.aip.org/apm/article/12/10/101110/3317172/Epitaxial-RuO2-and-IrO2-films-by-pulsed-laser},
   year = {2024}
}

@article{Lovesey2023,
   author = {S. W. Lovesey and D. D. Khalyavin and G. van der Laan},
   doi = {10.1103/PhysRevB.108.L121103},
   issn = {2469-9950},
   issue = {12},
   journal = {Physical Review B},
   month = {9},
   pages = {L121103},
   publisher = {American Physical Society},
   title = {{Magnetic structure of \ce{RuO2} in view of altermagnetism}},
   volume = {108},
   url = {https://link.aps.org/doi/10.1103/PhysRevB.108.L121103},
   year = {2023}
}

@article{Pawula2024,
   author = {Florent Pawula and Ali Fakih and Ramzy Daou and Sylvie Hébert and Natalia Mordvinova and Oleg Lebedev and Denis Pelloquin and Antoine Maignan},
   doi = {10.1103/PhysRevB.110.064432},
   issn = {2469-9950},
   issue = {6},
   journal = {Physical Review B},
   month = {8},
   pages = {064432},
   publisher = {American Physical Society},
   title = {{Multiband transport in \ce{RuO2}}},
   volume = {110},
   url = {https://link.aps.org/doi/10.1103/PhysRevB.110.064432},
   year = {2024}
}

@article{Brahimi2024,
   author = {Samy Brahimi and Dibya Prakash Rai and Samir Lounis},
   doi = {10.1088/1361-648X/ae084f},
   issn = {0953-8984},
   issue = {39},
   journal = {Journal of Physics: Condensed Matter},
   month = {9},
   pages = {395801},
   title = {{Confinement-induced altermagnetism in \ce{RuO2} thin films}},
   volume = {37},
   url = {https://iopscience.iop.org/article/10.1088/1361-648X/ae084f},
   year = {2025}
}

@article{Liu2024,
   author = {Jiayu Liu and Jie Zhan and Tongrui Li and Jishan Liu and Shufan Cheng and Yuming Shi and Liwei Deng and Meng Zhang and Chihao Li and Jianyang Ding and Qi Jiang and Mao Ye and Zhengtai Liu and Zhicheng Jiang and Siyu Wang and Qian Li and Yanwu Xie and Yilin Wang and Shan Qiao and Jinsheng Wen and Yan Sun and Dawei Shen},
   doi = {10.1103/PhysRevLett.133.176401},
   issn = {0031-9007},
   issue = {17},
   journal = {Physical Review Letters},
   month = {10},
   pages = {176401},
   title = {{Absence of Altermagnetic Spin Splitting Character in Rutile Oxide \ce{RuO2}}},
   volume = {133},
   url = {https://link.aps.org/doi/10.1103/PhysRevLett.133.176401},
   year = {2024}
}

@article{Jeong2024,
   author = {Seung Gyo Jeong and In Hyeok Choi and Sreejith Nair and Luca Buiarelli and Bita Pourbahari and Jin Young Oh and Bonnie Y.X. Lin and James M. LeBeau and Nabil Bassim and Daigorou Hirai and Ambrose Seo and Woo Seok Choi and Rafael M. Fernandes and Turan Birol and Liuyan Zhao and Jong Seok Lee and Bharat Jalan},
   doi = {10.1073/pnas.2526641123},
   issn = {0027-8424},
   issue = {10},
   journal = {Proceedings of the National Academy of Sciences},
   month = {3},
   title = {Altermagnetic polar metallic phase in ultrathin epitaxially strained RuO 2 films},
   volume = {123},
   url = {https://pnas.org/doi/10.1073/pnas.2526641123},
   year = {2026}
}

@article{Visscher2026,
   author = {Maria H. Visscher and Sebastian Buchberger and Bruno Saika and Shu Mo and Lea Richter and Mats Leandersson and Craig Polley and Andrew P. Mackenzie and Phil D. C. King},
   month = {5},
   title = {{Disentangling bulk and surface electronic structure using targeted cleave planes in \ce{RuO2}}},
   url = {http://arxiv.org/abs/2605.06798},
   year = {2026}
}

@article{Plouff2025,
   author = {David T. Plouff and Laura Scheuer and Shreya Shrestha and Weipeng Wu and Nawsher J. Parvez and Subhash Bhatt and Xinhao Wang and Lars Gundlach and M. Benjamin Jungfleisch and John Q. Xiao},
   doi = {10.1038/s44306-025-00083-2},
   issn = {2948-2119},
   issue = {1},
   journal = {npj Spintronics},
   month = {5},
   pages = {17},
   title = {{Revisiting altermagnetism in \ce{RuO2}: a study of laser-pulse induced charge dynamics by time-domain terahertz spectroscopy}},
   volume = {3},
   url = {https://www.nature.com/articles/s44306-025-00083-2},
   year = {2025}
}

@article{Peng2024,
   author = {Xin Peng and Zhihao Liu and Shengnan Zhang and Yi Zhou and Yuran Sun and Yahui Su and Chunxiang Wu and Tingyu Zhou and Le Liu and Yazhou Li and Hangdong Wang and Jinhu Yang and Bin Chen and Yuke Li and Chuanying Xi and Jianhua Du and Zhiwei Jiao and Quansheng Wu and Minghu Fang},
   doi = {10.1038/s43246-025-00905-0},
   issn = {2662-4443},
   issue = {1},
   journal = {Communications Materials},
   month = {8},
   pages = {177},
   title = {{Universal Scaling Behavior of Transport Properties in Non-Magnetic \ce{RuO2}}},
   volume = {6},
   url = {https://www.nature.com/articles/s43246-025-00905-0},
   year = {2025}
}

@article{Sukhachov2024,
   author = {Pavlo Sukhachov and Jacob Linder},
   doi = {10.1103/PhysRevB.110.205114},
   issn = {2469-9950},
   issue = {20},
   journal = {Physical Review B},
   month = {11},
   pages = {205114},
   title = {{Impurity-induced Friedel oscillations in altermagnets and $p$-wave magnets}},
   volume = {110},
   url = {https://link.aps.org/doi/10.1103/PhysRevB.110.205114},
   year = {2024}
}

@article{Kiefer2024,
   author = {L Kiefer and F Wirth and A Bertin and P Becker and L Bohatý and K Schmalzl and A Stunault and J A Rodríguez-Velamazan and O Fabelo and M Braden},
   doi = {10.1088/1361-648X/adad2a},
   issn = {0953-8984},
   issue = {13},
   journal = {Journal of Physics: Condensed Matter},
   month = {3},
   pages = {135801},
   pmid = {39898626},
   title = {{Crystal structure and absence of magnetic order in single-crystalline \ce{RuO2}}},
   volume = {37},
   url = {https://iopscience.iop.org/article/10.1088/1361-648X/adad2a},
   year = {2025}
}

@article{Ho2025,
  title = {{Symmetry-breaking induced surface magnetization in nonmagnetic \ce{RuO2}}},
  author = {Ho, Dai Q. and To, D. Quang and Hu, Ruiqi and Bryant, Garnett W. and Janotti, Anderson},
  journal = {Phys. Rev. Mater.},
  volume = {9},
  issue = {9},
  pages = {094406},
  numpages = {11},
  year = {2025},
  month = {Sep},
  publisher = {American Physical Society},
  doi = {10.1103/6fxv-153y},
  url = {https://link.aps.org/doi/10.1103/6fxv-153y}
}

@article{Hu2025,
   author = {Hao-Ran Hu and Xiangang Wan and Wei Chen},
   doi = {10.1103/PhysRevB.111.035132},
   issn = {2469-9950},
   issue = {3},
   journal = {Physical Review B},
   month = {1},
   pages = {035132},
   title = {{Quasiparticle interference in altermagnets}},
   volume = {111},
   url = {https://link.aps.org/doi/10.1103/PhysRevB.111.035132},
   year = {2025}
}

@article{Bloechl1994,
   author = {P. E. Blöchl},
   doi = {10.1103/PhysRevB.50.17953},
   isbn = {01631829/94/50(2},
   issn = {0163-1829},
   issue = {24},
   journal = {Physical Review B},
   month = {12},
   pages = {17953-17979},
   title = {Projector augmented-wave method},
   volume = {50},
   url = {https://link.aps.org/doi/10.1103/PhysRevB.50.17953},
   year = {1994}
}

@article{Kessler2024a,
   author = {Philipp Keßler and Laura Garcia-Gassull and Andreas Suter and Thomas Prokscha and Zaher Salman and Dmitry Khalyavin and Pascal Manuel and Fabio Orlandi and Igor I. Mazin and Roser Valentí and Simon Moser},
   doi = {10.1038/s44306-024-00055-y},
   issn = {2948-2119},
   issue = {1},
   journal = {npj Spintronics},
   month = {10},
   pages = {50},
   title = {{Absence of magnetic order in \ce{RuO2}: insights from $\mu$SR spectroscopy and neutron diffraction}},
   volume = {2},
   url = {https://www.nature.com/articles/s44306-024-00055-y},
   year = {2024}
}

@article{Guo2024,
   author = {Yaqin Guo and Jing Zhang and Zengtai Zhu and Yuan‐yuan Jiang and Longxing Jiang and Chuangwen Wu and Jing Dong and Xing Xu and Wenqing He and Bin He and Zhiheng Huang and Luojun Du and Guangyu Zhang and Kehui Wu and Xiufeng Han and Ding‐fu Shao and Guoqiang Yu and Hao Wu},
   doi = {10.1002/advs.202400967},
   issn = {2198-3844},
   issue = {25},
   journal = {Advanced Science},
   month = {7},
   pages = {e2400967},
   pmid = {38626379},
   publisher = {John Wiley and Sons Inc},
   title = {{Direct and Inverse Spin Splitting Effects in Altermagnetic \ce{RuO2}}},
   volume = {11},
   url = {https://advanced.onlinelibrary.wiley.com/doi/10.1002/advs.202400967},
   year = {2024}
}

@article{Zhou2024,
   author = {Xiaodong Zhou and Wanxiang Feng and Run-Wu Zhang and Libor Šmejkal and Jairo Sinova and Yuriy Mokrousov and Yugui Yao},
   doi = {10.1103/PhysRevLett.132.056701},
   issn = {0031-9007},
   issue = {5},
   journal = {Physical Review Letters},
   month = {1},
   pages = {056701},
   pmid = {38364129},
   publisher = {American Physical Society},
   title = {{Crystal Thermal Transport in Altermagnetic \ce{RuO2}}},
   volume = {132},
   url = {https://link.aps.org/doi/10.1103/PhysRevLett.132.056701},
   year = {2024}
}

@article{Over2016,
   author = {A. Goriachko and H. Over},
   doi = {10.1186/s11671-016-1757-2},
   issn = {1931-7573},
   issue = {1},
   journal = {Nanoscale Research Letters},
   month = {12},
   pages = {534},
   publisher = {Springer New York LLC},
   title = {{The Nanostructuring of Atomically Flat \ce{Ru(0001)} upon Oxidation and Reduction}},
   volume = {11},
   url = {https://link.springer.com/10.1186/s11671-016-1757-2},
   year = {2016}
}

@article{Schafer1963,
   author = {Harald Schäfer and Gerd Schneidereit and Wilfried Gerhardt},
   doi = {10.1002/zaac.19633190514},
   issn = {0044-2313},
   issue = {5-6},
   journal = {Zeitschrift für anorganische und allgemeine Chemie},
   month = {1},
   pages = {327-336},
   title = {{Zur Chemie der Platinmetalle. \ce{RuO2} Chemischer Transport, Eigenschaften, thermischer Zerfall}},
   volume = {319},
   url = {http://doi.wiley.com/10.1002/zaac.19633190514},
   year = {1963},
}

@article{Vadimsky1979,
   author = {R. G. Vadimsky and R. P. Frankenthal and D. E. Thompson},
   doi = {10.1149/1.2128846},
   issn = {0013-4651},
   issue = {11},
   journal = {Journal of The Electrochemical Society},
   month = {11},
   pages = {2017-2023},
   title = {{\ce{Ru} and \ce{RuO2} as Electrical Contact Materials: Preparation and Environmental Interactions}},
   volume = {126},
   url = {https://iopscience.iop.org/article/10.1149/1.2128846},
   year = {1979}
}

@article{Over2002a,
   author = {H. Over and A.P. Seitsonen and E. Lundgren and M. Schmid and P. Varga},
   doi = {10.1016/S0039-6028(02)01853-8},
   issn = {00396028},
   issue = {1},
   journal = {Surface Science},
   month = {8},
   pages = {143-156},
   title = {{Experimental and simulated STM images of stoichiometric and partially reduced \ce{RuO2(110)} surfaces including adsorbates}},
   volume = {515},
   url = {https://linkinghub.elsevier.com/retrieve/pii/S0039602802018538},
   year = {2002}
}

@article{Over2000,
   author = {H. Over and Y. D. Kim and A. P. Seitsonen and S. Wendt and E. Lundgren and M. Schmid and P. Varga and A. Morgante and G. Ertl},
   doi = {10.1126/science.287.5457.1474},
   issn = {0036-8075},
   issue = {5457},
   journal = {Science},
   month = {2},
   pages = {1474-1476},
   title = {{Atomic-Scale Structure and Catalytic Reactivity of the \ce{RuO2(110)} Surface}},
   volume = {287},
   url = {https://www.science.org/doi/10.1126/science.287.5457.1474},
   year = {2000}
}

@article{Jovic2019,
   author = {Vedran Jovic and Roland J. Koch and Swarup K. Panda and Helmuth Berger and Philippe Bugnon and Arnaud Magrez and Ronny Thomale and Kevin E. Smith and Silke Biermann and Chris Jozwiak and Aaron Bostwick and Eli Rotenberg and Domenico Di Sante and Simon Moser},
   journal = {cond-mat.mes-hall},
   month = {8},
   title = {{The Dirac nodal line network in non-symmorphic rutile semimetal \ce{RuO2}}},
   url = {http://arxiv.org/abs/1908.02621},
   year = {2019}
}

@article{Feng2022,
   author = {Zexin Feng and Xiaorong Zhou and Libor Šmejkal and Lei Wu and Zengwei Zhu and Huixin Guo and Rafael González-Hernández and Xiaoning Wang and Han Yan and Peixin Qin and Xin Zhang and Haojiang Wu and Hongyu Chen and Ziang Meng and Li Liu and Zhengcai Xia and Jairo Sinova and Tomáš Jungwirth and Zhiqi Liu},
   doi = {10.1038/s41928-022-00866-z},
   issn = {2520-1131},
   issue = {11},
   journal = {Nature Electronics},
   month = {11},
   pages = {735-743},
   title = {{An anomalous Hall effect in altermagnetic ruthenium dioxide}},
   volume = {5},
   url = {https://www.nature.com/articles/s41928-022-00866-z},
   year = {2022}
}

@article{Perdew1992,
   author = {John P. Perdew and J. A. Chevary and S. H. Vosko and Koblar A. Jackson and Mark R. Pederson and D. J. Singh and Carlos Fiolhais},
   doi = {10.1103/PhysRevB.46.6671},
   issn = {0163-1829},
   issue = {11},
   journal = {Physical Review B},
   month = {9},
   pages = {6671-6687},
   title = {Atoms, molecules, solids, and surfaces: Applications of the generalized gradient approximation for exchange and correlation},
   volume = {46},
   url = {https://link.aps.org/doi/10.1103/PhysRevB.46.6671},
   year = {1992}
}

@article{Over2012,
   author = {Herbert Over},
   doi = {10.1021/cr200247n},
   issn = {0009-2665},
   issue = {6},
   journal = {Chemical Reviews},
   month = {6},
   pages = {3356-3426},
   title = {{Surface Chemistry of Ruthenium Dioxide in Heterogeneous Catalysis and Electrocatalysis: From Fundamental to Applied Research}},
   volume = {112},
   url = {http://pubs.acs.org/doi/abs/10.1021/cr200247n},
   year = {2012},
}

@article{Jovic2018,
   author = {Vedran Jovic and Roland J. Koch and Swarup K. Panda and Helmuth Berger and Philippe Bugnon and Arnaud Magrez and Kevin E. Smith and Silke Biermann and Chris Jozwiak and Aaron Bostwick and Eli Rotenberg and Simon Moser},
   doi = {10.1103/PhysRevB.98.241101},
   issn = {2469-9950},
   issue = {24},
   journal = {Physical Review B},
   month = {12},
   pages = {241101},
   title = {{Dirac nodal lines and flat-band surface state in the functional oxide \ce{RuO2}}},
   volume = {98},
   url = {https://link.aps.org/doi/10.1103/PhysRevB.98.241101},
   year = {2018}
}

@article{Herd2012,
   author = {Benjamin Herd and Marcus Knapp and Herbert Over},
   doi = {10.1021/jp3085155},
   issn = {1932-7447},
   issue = {46},
   journal = {The Journal of Physical Chemistry C},
   month = {11},
   pages = {24649-24660},
   publisher = {American Chemical Society},
   title = {{Atomic Scale Insights into the Initial Oxidation of \ce{Ru(0001)} Using Molecular Oxygen: A Scanning Tunneling Microscopy Study}},
   volume = {116},
   url = {https://pubs.acs.org/doi/10.1021/jp3085155},
   year = {2012}
}

@article{Rossler2007a,
   author = {M. Rössler and S. Günther and J. Wintterlin},
   doi = {10.1021/jp065182j},
   issn = {1932-7447},
   issue = {5},
   journal = {The Journal of Physical Chemistry C},
   month = {2},
   pages = {2242-2250},
   title = {{Scanning Tunneling Microscopy of the \ce{RuO2(110)} Surface at Ambient Oxygen Pressure}},
   volume = {111},
   url = {https://pubs.acs.org/doi/10.1021/jp065182j},
   year = {2007}
}

@article{Sun2017,
   author = {Yan Sun and Yang Zhang and Chao-Xing Liu and Claudia Felser and Binghai Yan},
   doi = {10.1103/PhysRevB.95.235104},
   issn = {2469-9950},
   issue = {23},
   journal = {Physical Review B},
   month = {6},
   pages = {235104},
   title = {{Dirac nodal lines and induced spin Hall effect in metallic rutile oxides}},
   volume = {95},
   url = {http://link.aps.org/doi/10.1103/PhysRevB.95.235104},
   year = {2017}
}

@article{Zhu2019,
   author = {Z. H. Zhu and J. Strempfer and R. R. Rao and C. A. Occhialini and J. Pelliciari and Y. Choi and T. Kawaguchi and H. You and J. F. Mitchell and Y. Shao-Horn and R. Comin},
   doi = {10.1103/PhysRevLett.122.017202},
   issn = {0031-9007},
   issue = {1},
   journal = {Physical Review Letters},
   month = {1},
   pages = {017202},
   title = {{Anomalous Antiferromagnetism in Metallic \ce{RuO2} Determined by Resonant X-ray Scattering}},
   volume = {122},
   url = {https://link.aps.org/doi/10.1103/PhysRevLett.122.017202},
   year = {2019}
}

@article{Berlijn2017,
   author = {T. Berlijn and P. C. Snijders and O. Delaire and H.-D. Zhou and T. A. Maier and H.-B. Cao and S.-X. Chi and M. Matsuda and Y. Wang and M. R. Koehler and P. R. C. Kent and H. H. Weitering},
   doi = {10.1103/PhysRevLett.118.077201},
   issn = {0031-9007},
   issue = {7},
   journal = {Physical Review Letters},
   month = {2},
   pages = {077201},
   title = {{Itinerant Antiferromagnetism in \ce{RuO2}}},
   volume = {118},
   url = {https://link.aps.org/doi/10.1103/PhysRevLett.118.077201},
   year = {2017}
}

@article{Torun2013,
   author = {E. Torun and C. M. Fang and G. A. de Wijs and R. A. de Groot},
   doi = {10.1021/jp4020367},
   issn = {1932-7447},
   issue = {12},
   journal = {The Journal of Physical Chemistry C},
   month = {3},
   pages = {6353-6357},
   title = {{Role of Magnetism in Catalysis: \ce{RuO2(110)} Surface}},
   volume = {117},
   url = {https://pubs.acs.org/doi/10.1021/jp4020367},
   year = {2013}
}

@article{Kim2000,
   author = {Y.D. Kim and A.P. Seitsonen and H. Over},
   doi = {10.1016/S0039-6028(00)00733-0},
   issn = {00396028},
   issue = {1-2},
   journal = {Surface Science},
   month = {10},
   pages = {1-8},
   title = {{The atomic geometry of oxygen-rich \ce{Ru(0001)} surfaces: coexistence of (1×1)\ce{O} and \ce{RuO2(110)} domains}},
   volume = {465},
   url = {https://linkinghub.elsevier.com/retrieve/pii/S0039602800007330},
   year = {2000}
}

@article{He2015,
   author = {Yunbin He and Daniel Langsdorf and Lei Li and Herbert Over},
   doi = {10.1021/jp5121405},
   issn = {1932-7447},
   issue = {5},
   journal = {The Journal of Physical Chemistry C},
   month = {2},
   pages = {2692-2702},
   title = {{Versatile Model System for Studying Processes Ranging from Heterogeneous to Photocatalysis: Epitaxial \ce{RuO2(110)} on \ce{TiO2(110)}}},
   volume = {119},
   url = {https://pubs.acs.org/doi/10.1021/jp5121405},
   year = {2015}
}

@article{Kim2001,
   author = {Y. D. Kim and A. P. Seitsonen and S. Wendt and J. Wang and C. Fan and K. Jacobi and H. Over and G. Ertl},
   doi = {10.1021/jp003213j},
   issn = {1520-6106},
   issue = {18},
   journal = {The Journal of Physical Chemistry B},
   month = {5},
   pages = {3752-3758},
   title = {{Characterization of Various Oxygen Species on an Oxide Surface: \ce{RuO2(110)}}},
   volume = {105},
   url = {https://pubs.acs.org/doi/10.1021/jp003213j},
   year = {2001}
}

@article{Lee2012,
   author = {Youngmin Lee and Jin Suntivich and Kevin J. May and Erin E. Perry and Yang Shao-Horn},
   doi = {10.1021/jz2016507},
   isbn = {1948-7185},
   issn = {1948-7185},
   issue = {3},
   journal = {The Journal of Physical Chemistry Letters},
   month = {2},
   pages = {399-404},
   pmid = {22393313},
   title = {{Synthesis and Activities of Rutile \ce{IrO2} and \ce{RuO2} Nanoparticles for Oxygen Evolution in Acid and Alkaline Solutions}},
   volume = {3},
   url = {https://pubs.acs.org/doi/10.1021/jz2016507},
   year = {2012}
}

@article{Cotton1966,
   author = {F. A. Cotton and J. T. Mague},
   doi = {10.1021/ic50036a037},
   issn = {0020-1669},
   issue = {2},
   journal = {Inorganic Chemistry},
   month = {2},
   pages = {317-318},
   title = {{The Crystal and Molecular Structure of Tetragonal Ruthenium Dioxide}},
   volume = {5},
   url = {https://pubs.acs.org/doi/abs/10.1021/ic50036a037},
   year = {1966}
}

@article{Ahn2019,
   author = {Kyo-Hoon Ahn and Atsushi Hariki and Kwan-Woo Lee and Jan Kuneš},
   doi = {10.1103/PhysRevB.99.184432},
   issn = {2469-9950},
   issue = {18},
   journal = {Physical Review B},
   month = {5},
   pages = {184432},
   publisher = {American Physical Society},
   title = {{Antiferromagnetism in \ce{RuO2} as d-wave Pomeranchuk instability}},
   volume = {99},
   url = {https://link.aps.org/doi/10.1103/PhysRevB.99.184432},
   year = {2019}
}

@article{Lovesey2022,
   author = {S. W. Lovesey and D. D. Khalyavin and G. van der Laan},
   doi = {10.1103/PhysRevB.105.014403},
   issn = {2469-9950},
   issue = {1},
   journal = {Physical Review B},
   month = {1},
   pages = {014403},
   publisher = {American Physical Society},
   title = {{Magnetic properties of \ce{RuO2} and charge-magnetic interference in Bragg diffraction of circularly polarized x-rays}},
   volume = {105},
   url = {https://link.aps.org/doi/10.1103/PhysRevB.105.014403},
   year = {2022}
}

@article{Jovic2021,
   author = {Vedran Jovic and Armando Consiglio and Kevin E. Smith and Chris Jozwiak and Aaron Bostwick and Eli Rotenberg and Domenico Di Sante and Simon Moser},
   doi = {10.1021/acscatal.0c04871},
   issn = {2155-5435},
   issue = {3},
   journal = {ACS Catalysis},
   month = {2},
   pages = {1749-1757},
   title = {{Momentum for Catalysis: How Surface Reactions Shape the \ce{RuO2} Flat Surface State}},
   volume = {11},
   url = {https://pubs.acs.org/doi/10.1021/acscatal.0c04871},
   year = {2021}
}

@article{Over2004a,
   author = {H. Over and M. Knapp and E. Lundgren and A. P. Seitsonen and M. Schmid and P. Varga},
   doi = {10.1002/cphc.200300833},
   isbn = {6419934559},
   issn = {1439-4235},
   issue = {2},
   journal = {ChemPhysChem},
   month = {2},
   pages = {167-174},
   title = {{Visualization of Atomic Processes on Ruthenium Dioxide using Scanning Tunneling Microscopy}},
   volume = {5},
   url = {https://chemistry-europe.onlinelibrary.wiley.com/doi/10.1002/cphc.200300833},
   year = {2004}
}

@article{Mallet2001,
   author = {P. Mallet and H. Guyot and J. Y. Veuillen and N. Motta},
   doi = {10.1103/PhysRevB.63.165428},
   issn = {0163-1829},
   issue = {16},
   journal = {Physical Review B},
   month = {4},
   pages = {165428},
   title = {{Charge-density-wave STM observation in $\eta$–\ce{Mo4O11}}},
   volume = {63},
   url = {https://link.aps.org/doi/10.1103/PhysRevB.63.165428},
   year = {2001}
}

@article{Rodriguez1999,
   author = {P. Mallet and K. M. Zimmermann and Ph. Chevalier and J. Marcus and J. Y. Veuillen and J. M. Gomez Rodriguez},
   doi = {10.1103/PhysRevB.60.2122},
   issn = {0163-1829},
   issue = {3},
   journal = {Physical Review B},
   month = {7},
   pages = {2122-2126},
   title = {{Contrast reversal of the charge density wave STM image in purple potassium molybdenum bronze \ce{K_{0.9}Mo6O17}}},
   volume = {60},
   url = {https://link.aps.org/doi/10.1103/PhysRevB.60.2122},
   year = {1999}
}

@article{Smejkal2022a,
   author = {Libor Šmejkal and Jairo Sinova and Tomas Jungwirth},
   doi = {10.1103/PhysRevX.12.040501},
   issn = {2160-3308},
   issue = {4},
   journal = {Physical Review X},
   month = {12},
   pages = {040501},
   title = {Emerging Research Landscape of Altermagnetism},
   volume = {12},
   url = {https://link.aps.org/doi/10.1103/PhysRevX.12.040501},
   year = {2022}
}

@article{Spera2020,
   author = {M. Spera and A. Scarfato and A. Pasztor and E. Giannini and D. R. Bowler and Ch Renner},
   doi = {10.1103/PhysRevLett.125.267603},
   issn = {10797114},
   issue = {26},
   journal = {Physical Review Letters},
   pages = {267603},
   pmid = {33449793},
   publisher = {American Physical Society},
   title = {Insight into the Charge Density Wave Gap from Contrast Inversion in Topographic STM Images},
   volume = {125},
   url = {https://doi.org/10.1103/PhysRevLett.125.267603},
   year = {2020},
}

@article{Smejkal2022b,
   author = {Libor Šmejkal and Jairo Sinova and Tomas Jungwirth},
   doi = {10.1103/PhysRevX.12.031042},
   issn = {2160-3308},
   issue = {3},
   journal = {Physical Review X},
   month = {9},
   pages = {031042},
   publisher = {American Physical Society},
   title = {Beyond Conventional Ferromagnetism and Antiferromagnetism: A Phase with Nonrelativistic Spin and Crystal Rotation Symmetry},
   volume = {12},
   url = {https://link.aps.org/doi/10.1103/PhysRevX.12.031042},
   year = {2022}
}

@article{Kim2001a,
   author = {Y. D. Kim and S. Schwegmann and A. P. Seitsonen and H. Over},
   doi = {10.1021/jp003650y},
   issn = {1520-6106},
   issue = {11},
   journal = {The Journal of Physical Chemistry B},
   month = {3},
   pages = {2205-2211},
   title = {{Epitaxial Growth of \ce{RuO2(100)} on $\ce{Ru(10\bar{1} 0)}$: Surface Structure and Other Properties}},
   volume = {105},
   url = {https://pubs.acs.org/doi/10.1021/jp003650y},
   year = {2001}
}

@article{Tschirner2023,
   author = {Teresa Tschirner and Philipp Keßler and Ruben Dario Gonzalez Betancourt and Tommy Kotte and Dominik Kriegner and Bernd Büchner and Joseph Dufouleur and Martin Kamp and Vedran Jovic and Libor Smejkal and Jairo Sinova and Ralph Claessen and Tomas Jungwirth and Simon Moser and Helena Reichlova and Louis Veyrat},
   doi = {10.1063/5.0160335},
   issn = {2166-532X},
   issue = {10},
   journal = {APL Materials},
   month = {10},
   pages = {101103},
   publisher = {American Institute of Physics Inc.},
   title = {{Saturation of the anomalous Hall effect at high magnetic fields in altermagnetic \ce{RuO2}}},
   volume = {11},
   url = {https://pubs.aip.org/apm/article/11/10/101103/2913994/Saturation-of-the-anomalous-Hall-effect-at-high},
   year = {2023}
}

@article{Fedchenko2024,
   author = {Olena Fedchenko and Jan Minár and Akashdeep Akashdeep and Sunil Wilfred D’Souza and Dmitry Vasilyev and Olena Tkach and Lukas Odenbreit and Quynh Nguyen and Dmytro Kutnyakhov and Nils Wind and Lukas Wenthaus and Markus Scholz and Kai Rossnagel and Moritz Hoesch and Martin Aeschlimann and Benjamin Stadtmüller and Mathias Kläui and Gerd Schönhense and Tomas Jungwirth and Anna Birk Hellenes and Gerhard Jakob and Libor Šmejkal and Jairo Sinova and Hans-Joachim Elmers},
   doi = {10.1126/sciadv.adj4883},
   issn = {2375-2548},
   issue = {5},
   journal = {Science Advances},
   month = {2},
   pages = {31},
   title = {{Observation of time-reversal symmetry breaking in the band structure of altermagnetic \ce{RuO2}}},
   volume = {10},
   url = {https://www.science.org/doi/10.1126/sciadv.adj4883},
   year = {2024}
}

@techReport{Swanson1955,
   author = {Howard Swanson and Ruth Fuyat and George Ugrinic},
   institution = {National Bureau of Standards},
   month = {3},
   pages = {5-6},
   title = {{Standard X-ray Diffraction Powder Patterns, NBS CIRCULAR 539, Volume IV}},
   url = {https://nvlpubs.nist.gov/nistpubs/Legacy/circ/nbscircular539v4.pdf},
   year = {1955}
}

@article{White_2005,
   author = {Steven R. White},
   doi = {10.1103/PhysRevB.72.180403},
   issn = {1098-0121},
   issue = {18},
   journal = {Physical Review B},
   month = {11},
   pages = {180403},
   title = {{Density matrix renormalization group algorithms with a single center site}},
   volume = {72},
   url = {https://link.aps.org/doi/10.1103/PhysRevB.72.180403},
   year = {2005}
}

@article{SCHOLLWOCK201196,
title = {{The density-matrix renormalization group in the age of matrix product states}},
journal = {Annals of Physics},
volume = {326},
number = {1},
pages = {96-192},
year = {2011},
note = {January 2011 Special Issue},
issn = {0003-4916},
doi = {https://doi.org/10.1016/j.aop.2010.09.012},
url = {https://www.sciencedirect.com/science/article/pii/S0003491610001752},
author = {Ulrich Schollwöck}
}

@article{ITensor,
	title={{The ITensor Software Library for Tensor Network Calculations}},
	author={Matthew Fishman and Steven R. White and E. Miles Stoudenmire},
	journal={SciPost Phys. Codebases},
	pages={4},
	year={2022},
	publisher={SciPost},
	doi={10.21468/SciPostPhysCodeb.4},
	url={https://scipost.org/10.21468/SciPostPhysCodeb.4},
}

@article{White_1992,
  title = {{Density matrix formulation for quantum renormalization groups}},
  author = {White, Steven R.},
  journal = {Phys. Rev. Lett.},
  volume = {69},
  issue = {19},
  pages = {2863--2866},
  numpages = {0},
  year = {1992},
  month = {Nov},
  publisher = {American Physical Society},
  doi = {10.1103/PhysRevLett.69.2863},
  url = {https://link.aps.org/doi/10.1103/PhysRevLett.69.2863}
}

\end{document}